\documentclass[conference,letterpaper,onecolumn,11pt]{IEEEtran}
\IEEEoverridecommandlockouts

\usepackage{cite}
\usepackage{amsmath,amssymb,amsfonts}
\usepackage{bbm}
\usepackage{algorithmic}
\usepackage{array}
\usepackage{graphicx}
\usepackage{textcomp}
\usepackage{xcolor}
\usepackage{tikz}
\usepackage{booktabs}
\usepackage{tabularx}
\usepackage{placeins}
\newcommand\copyrighttext{%
  \footnotesize \textcopyright{} 2026 IEEE. Personal use of this material is permitted.
  Permission from IEEE must be obtained for all other uses, in any current or future
  media, including reprinting/republishing this material for advertising or promotional
  purposes, creating new collective works, for resale or redistribution to servers or
  lists, or reuse of any copyrighted component of this work in other works.}
\newcommand\copyrightnotice{%
  \begin{tikzpicture}[remember picture,overlay]
    \node[anchor=south,yshift=10pt] at (current page.south)
    {\fbox{\parbox{\dimexpr\textwidth-2\fboxsep-2\fboxrule\relax}{\copyrighttext}}};
  \end{tikzpicture}%
}
\def\BibTeX{{\rm B\kern-.05em{\sc i\kern-.025em b}\kern-.08em
  T\kern-.1667em\lower.7ex\hbox{E}\kern-.125emX}}
\begin{document}

\title{ML-OLLA: Link Adaptation for Real-Time Traffic in 5G/6G NTN Systems
\thanks{The research has been supported by the Russian Science Foundation (Grant No 25-19-00659, https://rscf.ru/project/25-19-00659/)}
}

\author{\IEEEauthorblockN{
Artur Magamedov,
Artem Krasilov,
Roman Zlobin, and
Evgeny Khorov}
\IEEEauthorblockA{HSE University, Moscow, Russia}
\IEEEauthorblockA{Institute for Information Transmission Problems of the Russian Academy of Sciences, Moscow, Russia}
\IEEEauthorblockA{\{magamedov, krasilov, zlobin, khorov\}@wnlab.ru}
}

\maketitle
\copyrightnotice

\begin{abstract}
Reliable low-latency connectivity for real-time applications is a key requirement for 5G and future 6G non-terrestrial networks. Low Earth Orbit satellites enable global coverage but impose long propagation delays, rapid channel variations, and delayed feedback, which challenge conventional link adaptation algorithms designed for terrestrial networks. To address this problem, the paper proposes a Multi-Level Outer-Loop Link Adaptation (ML-OLLA) algorithm that assigns a dedicated target block error rate and an independent adaptation process for each transmission attempt. With ML-OLLA, early transmission attempts use higher-rate modulation and coding schemes (MCSs), while later ones use more robust MCSs. The core contribution of this work is a low-complexity data-driven parameter selection method that determines ML-OLLA parameters. Numerical results obtained with ns-3 show that the proposed ML-OLLA algorithm increases the system capacity by up to four times and reduces the minimum supported latency budget. In particular, for the 99.9\% reliability target, ML-OLLA reduces the minimum supported latency budget from 60~ms to 40~ms compared with the baseline OLLA.
\end{abstract}

\begin{IEEEkeywords}
 5G, 6G, LEO NTN, link adaptation, OLLA, real-time traffic
\end{IEEEkeywords}

\section{Introduction}
The widespread adoption of fifth-generation (5G) cellular systems has brought significant performance improvements in terms of throughput, latency, and reliability compared to 4G systems. Nevertheless, due to the high infrastructure costs of Terrestrial Networks (TN), coverage remains concentrated in urban and suburban areas, leaving rural regions, transport corridors, and remote locations underserved or entirely without 5G connectivity. To overcome this limitation, in Release 17, the 3rd Generation Partnership Project (3GPP) standardized Non-Terrestrial Networks (NTN), enabling the use of satellites or high-altitude platforms for providing cellular connectivity. By leveraging satellite constellations, operators can provide broadband access to virtually any point on the globe.

Most existing research on 5G NTN focuses on elastic data traffic (HTTP, FTP, video-on-demand), which tolerates the high latency inherent to satellite links. In contrast, we consider real-time traffic relevant to both user services and remote-area services, including voice and video communication, AR/VR/XR, real-time gaming, mission-critical communication, drones, and intelligent transport systems. Such applications typically require end-to-end packet latency of the order of tens of milliseconds, as detailed in Table~\ref{tab1}. Such stringent latency requirements are attainable only with Low Earth Orbit (LEO) satellites or high-altitude platforms.

\begin{table}[!htbp]
  \caption{Examples of real-time applications}
  \label{tab1}
  \centering
  \small
  \renewcommand{\arraystretch}{1.25}
  \setlength{\tabcolsep}{9pt}
  \begin{tabularx}{0.96\linewidth}{@{}>{\raggedright\arraybackslash}X >{\centering\arraybackslash}p{0.23\linewidth} >{\centering\arraybackslash}p{0.25\linewidth}@{}}
    \toprule
    \textbf{Application} & \textbf{Packet Delay Budget, ms} & \textbf{Maximum Packet Loss} \\
    \midrule
    Conversational voice & $100$ & $10^{-2}$ \\
    Mission Critical Push-To-Talk & $75$ & $10^{-2}$ \\
    Mission Critical Video & $100$ & $10^{-3}$ \\
    Live Streaming & $100\ldots150$ & $10^{-3}\ldots10^{-6}$ \\
    Real-Time Gaming & $50\ldots100$ & $10^{-3}\ldots10^{-4}$ \\
    AR/VR/XR & $5\ldots20$ & $10^{-3}\ldots10^{-6}$ \\
    Intelligent transport systems & $3\ldots50$ & $10^{-2}\ldots10^{-6}$ \\
    \bottomrule
  \end{tabularx}
\end{table}

Looking toward 6G~\cite{Khorov2026Current6G}, NTN is expected to evolve from a complementary extension into a fully integrated network component with massive LEO constellations and optical inter-satellite links. This evolution further strengthens the need for radio-access solutions that provide highly reliable and low-latency data transmission over the satellite link.

An important radio-access mechanism for satisfying strict Quality of Service (QoS) requirements for real-time traffic is Link Adaptation (LA). LA selects the modulation and coding scheme (MCS) for each transmission attempt and determines how many attempts can fit within the application's delay budget. In contrast to 5G TN systems, LA for 5G NTN systems faces the following challenges.
First, long propagation delays consume a large portion of the delay budget, leaving little room for retransmissions. Second, satellite and user mobility cause rapid channel variations, making accurate MCS selection difficult. Third, the feedback latency for channel quality reports and positive/negative acknowledgments (ACK/NACKs) is high with respect to the channel variation rate.

In this paper, we develop a novel LA algorithm for real-time traffic in 5G/6G NTN systems. The main idea of the proposed solution, called Multi-Level Outer-Loop Link Adaptation (ML-OLLA), is to assign different reliability targets and MCSs for various transmission attempts. The key contribution is a low-complexity data-driven parameter selection method that learns the per-attempt target error rates from the empirical distribution of observed MCS indexes. 
The results obtained with the ns-3 simulator show that the proposed ML-OLLA algorithm significantly increases the system capacity compared with the state-of-the-art LA algorithms designed for 5G TN systems. Moreover, ML-OLLA satisfies the strict latency requirements while consuming fewer radio resources than the existing LA algorithms. For the 99.9\% reliability target, ML-OLLA reduces the minimum supported latency budget from 60~ms to 40~ms compared with the baselines.

The rest of the paper is organized as follows. Section~\ref{sec:related-works} reviews LA algorithms for TN and NTN systems. Section~\ref{sec:scenario} describes the scenario and problem statement. Section~\ref{sec:prop-la} presents ML-OLLA algorithm while Section~\ref{sec:performance-evaluation} compares ML-OLLA with the state-of-the-art solutions. Section~\ref{sec:conclusion} concludes the paper.

\section{Related Works}
\label{sec:related-works}

Most existing studies on 5G NTN systems focus on elastic traffic and throughput-oriented performance evaluation. Specifically,~\cite{badini-2022-ns3,sedin-2020-leo} provide system-level 5G NTN simulation tools and evaluate the capacity of 5G NTN LEO systems  under simple traffic models, such as full-buffer traffic, using throughput as the primary metric. These studies do not consider real-time applications with specific Packet Loss Rate (PLR) and latency requirements. The authors of~\cite{tuninato-2025-harq} study retransmission mechanisms: Hybrid Automatic Repeat reQuest (HARQ) and Radio Link Control ARQ in terms of Block Error Rate (BLER), spectral efficiency, and latency. Unlike~\cite{badini-2022-ns3,sedin-2020-leo}, the study in~\cite{tuninato-2025-harq} considers the reliability and latency effects of retransmissions but does not consider the problem of LA under application-specific QoS constraints. This gap is especially important in 5G NTN LEO systems because delayed CQI reports, delayed HARQ feedback, and a small number of feasible transmission attempts within a packet latency budget make each parameter selection decision critical.

Since LA in 5G NTN LEO systems is an open problem, let us consider LA algorithms developed for 5G TNs. One group of works ~\cite{park2015optimizing,horita2024dynamic} proposes solutions that adaptively select the target BLER. Specifically, the paper~\cite{park2015optimizing} shows that using the same  target BLER for different transmission attempts is not always throughput-optimal, and more aggressive target BLERs for early attempts can improve spectral efficiency, while the paper~\cite{horita2024dynamic} shows that dynamic target BLER control can also improve TCP throughput under channel fluctuations. Another group of works~\cite{saxena2022reinforcement,wiesmayr2025salad} proposes solutions based on machine learning: reinforcement learning, Thompson sampling, and self-adaptive LA to select BLER targets from ACK/NACK feedback. The solutions listed above mainly optimize terrestrial throughput or long-term BLER behavior and do not determine reliability targets for different HARQ attempts under long delays over satellite links. Moreover, learning-based methods that require model inference or iterative training may be infeasible for satellite gNBs with limited onboard processing resources.

Many LA solutions for 5G TN systems are based on OLLA --- the most widely used algorithm that employs ACK/NACK feedback. Several works focus on improving its convergence time to the given reliability target. In particular, enhanced OLLA~\cite{blanquezcasado2016eolla} adaptively modifies the step size of OLLA and can update the offset even when no data packets are received. Non-linear OLLA~\cite{zhu2023nolla} uses an exponentially decaying adaptive step to reduce effective Signal-to-Interference-plus-Noise Ratio (SINR) fluctuations, while exponential-decay OLLA~\cite{mazumdar2026enhancing} combines instantaneous user-side channel measurements with long-term statistics to accelerate convergence. Low-BLER analysis further shows that stringent targets make NACK feedback sparse and less informative~\cite{chen2023effect}. These works improve convergence, stability, or feedback utilization, but they do not address how the target BLERs should be selected for the initial transmission and HARQ retransmissions.

The usage of different targets is useful because the resource cost of transmission attempts is asymmetric: the first transmission is always performed, whereas retransmission resources are consumed only after a previous failure. This trade-off is demonstrated for multicast URLLC traffic in~\cite{krasilov2024resource,Karamyshev2025Tutorial,Karamyshev2024Closed}, where different target BLERs reduce channel resource consumption compared with equal targets. However, this approach is limited to two attempts and uses fixed target BLER values. This work significantly extends the idea from~\cite{krasilov2024resource}: (i) we consider several HARQ retransmissions, (ii) we develop a method that adaptively selects target BLERs for all transmission attempts.

\section{Scenario and Problem Statement}
\label{sec:scenario}

\begin{figure}[!htbp]
  \centering
  \includegraphics[width=0.70\linewidth]{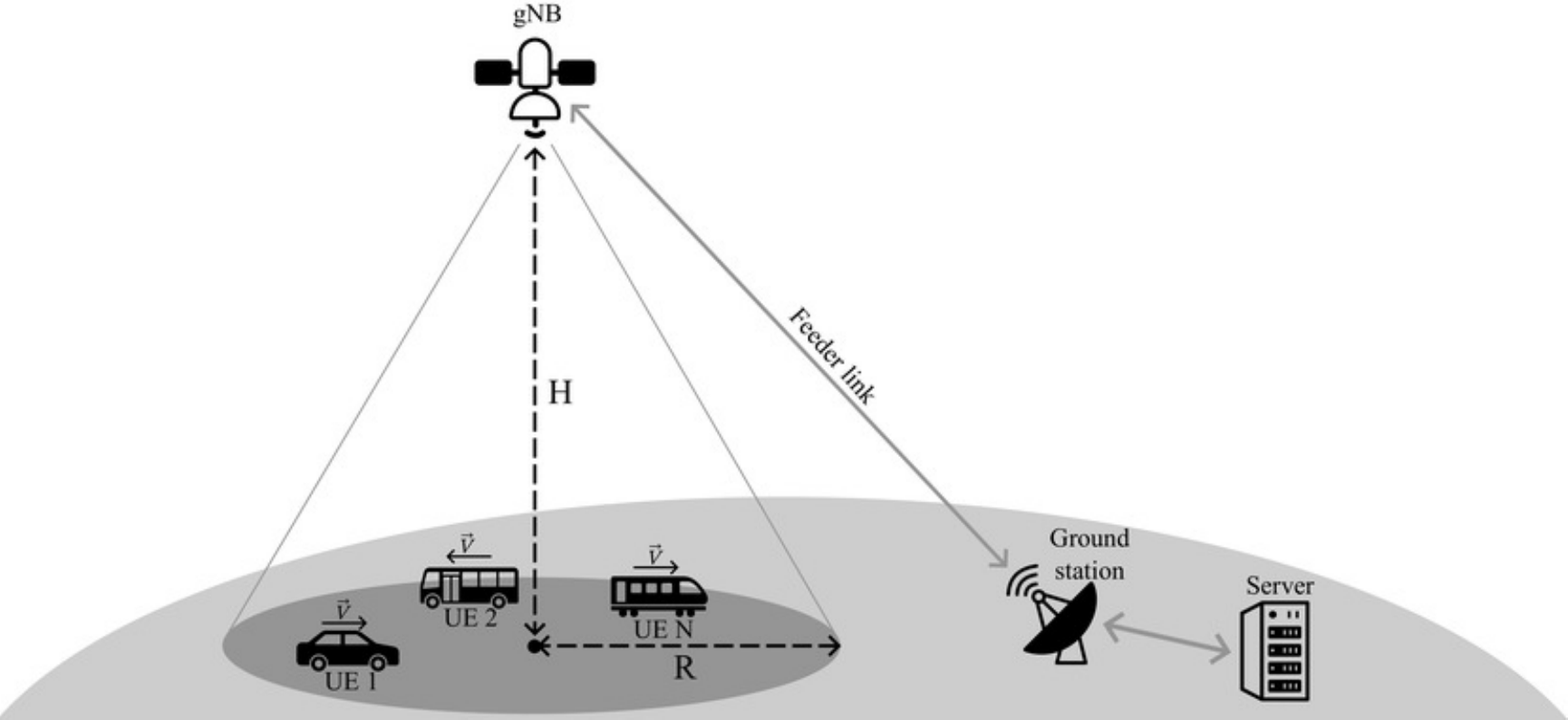}
  \caption{Considered LEO NTN scenario.}
  \label{fig:scenario-scheme}
\end{figure}

We consider a downlink 5G/6G LEO NTN scenario shown in Fig.~\ref{fig:scenario-scheme}. A LEO satellite provides downlink access to ground User Equipment (UEs) moving at different speeds. The next generation NodeB (gNB) functionality can be placed either on the ground or on board the satellite.
Using a directional antenna, the satellite forms a single spot beam that covers a circular area where mobile UEs are randomly distributed. The considered scenario focuses on a single serving beam. In the case of multi-beam satellites, we assume that inter-beam interference is reduced via frequency reuse schemes~\cite{tr38.821}.

If the gNB is located on the satellite, it is connected to the ground station through a feeder link, e.g., a Ka-band or optical feeder link. 
We assume that this feeder link does not limit the system performance: its capacity is much higher than the aggregate traffic of all UEs, while the propagation latency is determined only by the satellite and ground-station coordinates. 
The ground station is connected to the remote server through a high-capacity wired link with negligible delay. Thus, in the paper, we focus on analyzing latency in the NTN system.

The remote server generates real-time downlink data flows. We focus on downlink because the gNB directly controls downlink scheduling and MCS selection. The uplink case involves additional grant acquisition and UE-side buffer constraints and is left for future work. 
Each flow has two QoS requirements: the packet delivery time shall not exceed $D^{QoS}$, and the fraction of packets lost or delivered after this deadline shall not exceed $PLR^{QoS}$.

\textbf{Problem Statement.}
We define the system capacity $N_{\max}(\alpha)$ as the maximum number of simultaneously served UEs for which at least an $\alpha$ fraction of UEs satisfy both QoS constraints $D^{QoS}$ and $PLR^{QoS}$. Since all UEs within the considered beam share the same time-frequency resources, reducing the average number of Resource Blocks (RBs) consumed by each UE directly increases the number of real-time flows that can be supported under the same QoS constraints. Therefore, the goal of this paper is to design an LA algorithm that minimizes the channel resource consumption of each UE while satisfying the given QoS constraints. 

\section{Proposed LA Algorithm}
\label{sec:prop-la}

Consider a gNB that transmits a downlink data flow to a particular UE. We assume HARQ with Chase Combining (HARQ-CC) scheme for retransmissions. Given the latency budget $D^{QoS}$ and the HARQ Round-Trip Time (RTT) $T_{HARQ}$, which depends on satellite-link propagation delay and the processing capabilities of the gNB and UE, the gNB can estimate the maximum number of transmission attempts
$N=\lfloor D^{QoS} / T_{HARQ} \rfloor$ for each packet. For the considered LEO altitude and real-time delay budgets, $N$ is typically small. For example, $D^{QoS}=30$~ms allows only two transmission attempts in the gNB on ground scenario considered in Section~\ref{sec:performance-evaluation}. The LA algorithm aims to select the $MCS_n$ for each transmission attempt $n = 1, \ldots, N$, where $n = 1$ corresponds to the initial data transmission and $n>1$ corresponds to HARQ retransmissions.

The proposed MCS selection algorithm consists of three components: (i) Inner-Loop Link Adaptation (ILLA), described in Subsection~\ref{sec:illa}, which selects the MCS based on the latest channel measurements; (ii) ML-OLLA, described in Subsection~\ref{sec:ml-olla}, which corrects the MCS based on per-attempt ACK/NACK feedback; and (iii) the data-driven ML-OLLA parameter selection method described in Subsection~\ref{sec:optimization}, which learns per-MCS error probabilities from a sliding window of observed MCS indexes and determines the target BLERs and ML-OLLA step sizes.

\subsection{Inner-Loop Link Adaptation}
\label{sec:illa}
ILLA is a basic LA algorithm. It aims to find an MCS that fits the current channel conditions. We assume that each UE reports subband Channel Quality Indicators (CQIs) with period $T_{CQI}$. CQIs are integer values that correspond to quantized SINR measured at the UE in a given frequency band. More precisely, at time $t$, for each RB $b$, the gNB has the latest received CQI report: $CQI(t)=\{CQI_b(t),b=1,\ldots,B\}$, where $B$ is the total number of RBs.

When the gNB allocates a specific set $A(t)$ of RBs, ILLA finds the MCS $MCS^{ILLA}(t)=F(CQI(t),A(t))$ as a function of the measured CQIs in two steps. First, it maps CQI to MCS indexes $MCS(CQI_b(t))$ according to the CQI and MCS tables defined in 3GPP TS~38.214 (i.e., it finds the MCS and CQI indexes with the closest spectral efficiencies). Second, the gNB averages the MCS indexes over the set of allocated RBs $A(t)$ and rounds the resulting value to the closest integer MCS index $MCS^{ILLA}(t)$.

\subsection{ML-OLLA}
\label{sec:ml-olla}

ILLA faces significant challenges across diverse scenarios due to CQI quantization~\cite{Endovitskiy2026Channel}, the time-varying wireless channel, and the resulting CQI aging. Propagation delays and UE/satellite mobility misalign channel measurements with the current channel conditions, so the selected MCS may not meet the required reliability.

One of the ways to take into account channel variability is to correct ILLA decisions based on the ACK/NACK feedback provided by the UE during transmission. Existing algorithms --- the most widely known being OLLA --- consider only the selection of $MCS_1$ for the initial data transmission to provide the given target BLER $p_1$, while HARQ retransmissions are performed with the same MCS:
$MCS_n = MCS_1$ for $n \geq 2$.

The key idea of the proposed solution called ML-OLLA is to
assign different MCSs and reliability targets for various transmission attempts. Existing works show that to achieve lower resource consumption compared with OLLA, the target BLER for each retransmission shall be lower than that for the initial transmission: $p_1 \geq p_2 \geq \cdots \geq p_N$, such that
$\prod_{n=1}^{N} p_n \leq PLR^{QoS}$. Note that 3GPP specifications allow the use of different MCSs for HARQ retransmissions. However, the existing works do not describe how to select the MCS for each transmission attempt $n$ in order to achieve the given target BLERs $p_n$. To achieve this goal, we propose to run multiple OLLA processes at gNB. Specifically, for each transmission attempt $n$, the gNB keeps a real number $MCS^{sub}_n$, which provides a correction to the MCS selected based on the channel measurements (i.e., based on the ILLA algorithm). The MCS for transmission attempt $n$ is selected as $MCS_n=\operatorname{round}(MCS^{ILLA}_n-MCS^{sub}_n)$ and limited to the valid interval $[0,MCS_{\max}]$, where $MCS_{\max}$ is the maximum MCS index.

For HARQ retransmissions, ILLA re-evaluates the MCS using the latest CQI while enforcing $MCS_n \leq MCS_{n-1}$ for $n\geq2$, so that retransmissions do not use a higher MCS, reducing the risk of further transmission failures.

The MCS correction $MCS^{sub}_n$ takes into account the inaccuracies in ILLA MCS selection caused by measurement delay and CQI quantization. To maintain a given target BLER for transmission attempt $n$, $MCS^{sub}_n$ is updated based only on the ACK/NACKs received as feedback for attempt $n$. This approach allows different target BLERs to be maintained for different transmission attempts. Specifically, for each received ACK:
$MCS^{sub}_n = MCS^{sub}_n -
\Delta^{ACK}_n$, and for each received NACK:
$MCS^{sub}_n = MCS^{sub}_n +
\Delta^{NACK}_n$, where $\Delta^{ACK}_n$ and
$\Delta^{NACK}_n$ are the algorithm parameters. To maintain the
target BLER $p_n$ for transmission attempt $n$, the parameters shall satisfy the condition
$p_n = \frac{\Delta^{ACK}_n}
{\Delta^{ACK}_n + \Delta^{NACK}_n}$. In summary, each OLLA
process independently considers the statistics of successes and failures of the corresponding attempt to determine the required subtraction $MCS^{sub}_n$.
Note that each ML-OLLA process inherits the established in~\cite{blanquezcasado2016eolla} convergence properties of baseline OLLA.

The remaining problem is adaptive selection of $p_n$, $\Delta^{ACK}_n$, and $\Delta^{NACK}_n$.

\subsection{Parameter Selection}
\label{sec:optimization}

The main idea behind the proposed parameter selection method is to estimate the data transmission error probability for each MCS index $M$ ($M\in\{0,\ldots,M_{\max}\}$). This probability is not assumed to be time-invariant. Instead, at each update time $t$, the gNB computes a sliding-window estimate $\hat{\pi}_t(M)$ from recently received CQI reports. For each UE, the gNB keeps a window of $W$ consecutive CQI reports. When the gNB receives a CQI report, it runs a virtual allocation of RBs for a packet of a given size $L$. Specifically, it allocates RBs with the highest CQI values one by one and selects MCS using the ILLA algorithm described in Section~\ref{sec:illa}. As a result, for each report, the gNB obtains a sample of the MCS index that would be supported by the reported channel state. Let $f_{t,M}$ be the number of samples corresponding to MCS index $M$ in the current window of size $W$ by the time $t$. The window-based estimate of the error probability is given by $\hat{\pi}_t(M)=\sum_{M'< M}f_{t,M'}/W$. The cumulative frequency below $M$ is therefore used as a natural estimate for the probability that channel observations support an MCS lower than $M$.

Having the window-based estimates $\hat{\pi}_t(M)$, the parameter selection problem is formulated as a resource minimization problem. Specifically, let $M_n$ be the MCS index used for transmission attempt $n$ and $B(M_n,L)$ be the number of RBs required to transmit a packet of size $L$ using MCS index $M_n$. The optimization problem is to find the MCS indices for all transmission attempts that minimize the average resource consumption under the given reliability constraint: 

\begin{equation}
  \begin{aligned}
    &\underset{M_1,\dots,M_N}{\operatorname{minimize}}
    && \mathbb{E}[B]=\sum_{n=1}^{N} \left( \prod_{k=1}^{n-1} \hat{\pi}_t(M_{k}) \right) B(M_n, L) \\
    &\text{subject to}
    && \prod_{n=1}^{N} \hat{\pi}_t(M_n) \;\leq\; PLR^{QoS}.
  \end{aligned}
  \label{eq:optimization-problem}
\end{equation}

The expectation in~\eqref{eq:optimization-problem} is taken over all transmission attempts: attempt $n$ consumes resources only if all previous attempts fail. The parameter-selection model approximates decoding errors across attempts as independent and omits the HARQ-CC gain, thus conservatively overestimating packet-loss probability. Accounting for HARQ-CC is left for future work.

The discrete optimization uses exhaustive search over MCS combinations satisfying $M_n\leq M_{n-1}$ for $n\geq2$. Complexity and memory overhead remain low because the MCS set is finite and for each UE only the sliding MCS window, empirical CDF, and OLLA state are stored.

Having found the optimal values $M_n^{\star}$, the target BLER for OLLA process number $n$ is set to $\hat{\pi}_t(M_n^{\star})$. After that, OLLA steps $\Delta^{ACK}_n$ and $\Delta^{NACK}_n$ are derived. By default, we set $\Delta^{NACK}_n=1$ and determine $\Delta^{ACK}_n=\hat{\pi}_t(M_n^{\star})/(1-\hat{\pi}_t(M_n^{\star}))$.

\section{Performance Evaluation}
\label{sec:performance-evaluation}

\subsection{Simulation Setup}

We evaluate the scenario described in Section~\ref{sec:scenario} in ns-3~\cite{ns-3} with 5G-LENA module~\cite{5g-lena}, NTN propagation delays, and the 3GPP NTN channel model defined in TR~38.811~\cite{tr38.811}. Table~\ref{tab:leo_params} lists the main simulation parameters.

The gNB transmits Channel State Information Reference Signal (CSI-RS) with the period $T_{CSI-RS}$ and each UE reports CQIs with the period $T_{CQI}$. We explicitly model delayed CQI and HARQ feedback: the scheduler does not use the instantaneous channel state, but only the CQI reports and ACK/NACKs that have arrived at the gNB by the current time taking into account propagation delays.

We consider an Intelligent Transport Systems use case with the following traffic: for each UE, the server generates packets with the size of $L=256$ bytes and the period of 20~ms. As detailed in Section~\ref{sec:scenario}, we consider two gNB deployment options: with the gNB on the ground and gNB on the satellite.

We evaluate the following baselines. First, we consider conventional OLLA with the manual setting of the target BLER for the first transmission attempt. Specifically, if the delay budget allows $N$ transmission attempts and the target PLR after all transmission attempts is $P_{target}$, the target BLER $p_1=(P_{target})^{1/N}$. Note that typically $P_{target}$ is set to $PLR^{QoS}$. 
Second, we consider OLLA with the first-transmission target BLER adaptively selected for each UE using the SINR-based formula $p_1=f(SINR)$ from~\cite{park2015optimizing}. Third, the gNB always selects the most robust MCS. 

The proposed ML-OLLA algorithm is configured as follows: (i) the window $W$ is set to 2000 samples, (ii) the number of transmission attempts $N$ is determined based on the delay budget $D^{QoS}$, (iii) the target reliability $P_{target}=PLR^{QoS}$. The value $W=2000$ is chosen as a trade-off between estimation precision and adaptation speed: smaller windows react faster but produce noisier tail estimates, while larger windows smooth channel variations but adapt more slowly to channel changes.

In the experiments, we measure the following performance indicators: (i) the fraction of UEs for which QoS requirements are satisfied, (ii) the capacity defined as the maximal number of UEs for which at least $\alpha=95\%$ of UEs are satisfied.
All results are averaged over 15 independent simulation runs.

\begin{table}[!htbp]
  \caption{Main simulation parameters}
  \label{tab:leo_params}
  \centering
  \footnotesize
  \renewcommand{\arraystretch}{1.32}
  \setlength{\tabcolsep}{5pt}
  \begin{tabularx}{0.99\linewidth}{@{}>{\raggedright\arraybackslash}p{0.24\linewidth} >{\raggedright\arraybackslash}p{0.235\linewidth} >{\raggedright\arraybackslash}p{0.245\linewidth} >{\raggedright\arraybackslash}X@{}}
    \toprule
    \textbf{Parameter} & \textbf{Value} & \textbf{Parameter} & \textbf{Value} \\
    \midrule
    Satellite altitude $H$ & 600 km & UE antenna gain & 3 dBi \\
    Central frequency & 3.6 GHz & UE noise figure & 5 dB \\
    Channel bandwidth & 10 MHz & Beam footprint diameter & 50 km \\
    Satellite Tx power & 47 dBm & UE speed & 3...100 km/h \\
    gNB antenna aperture & 2 m & CSI-RS/CQI report period & 5 ms \\
    Satellite Tx max gain & 30 dBi & UE antenna model & IsotropicAntennaModel \\
    \multicolumn{2}{@{}l}{Satellite antenna model} & \multicolumn{2}{l@{}}{CircularApertureAntennaModel} \\
    \bottomrule
  \end{tabularx}
\end{table}

\subsection{Analysis of Results}

\begin{figure}[!htbp]
  \centering
  \begin{minipage}[t]{0.485\linewidth}
    \centering
    \includegraphics[width=\linewidth]{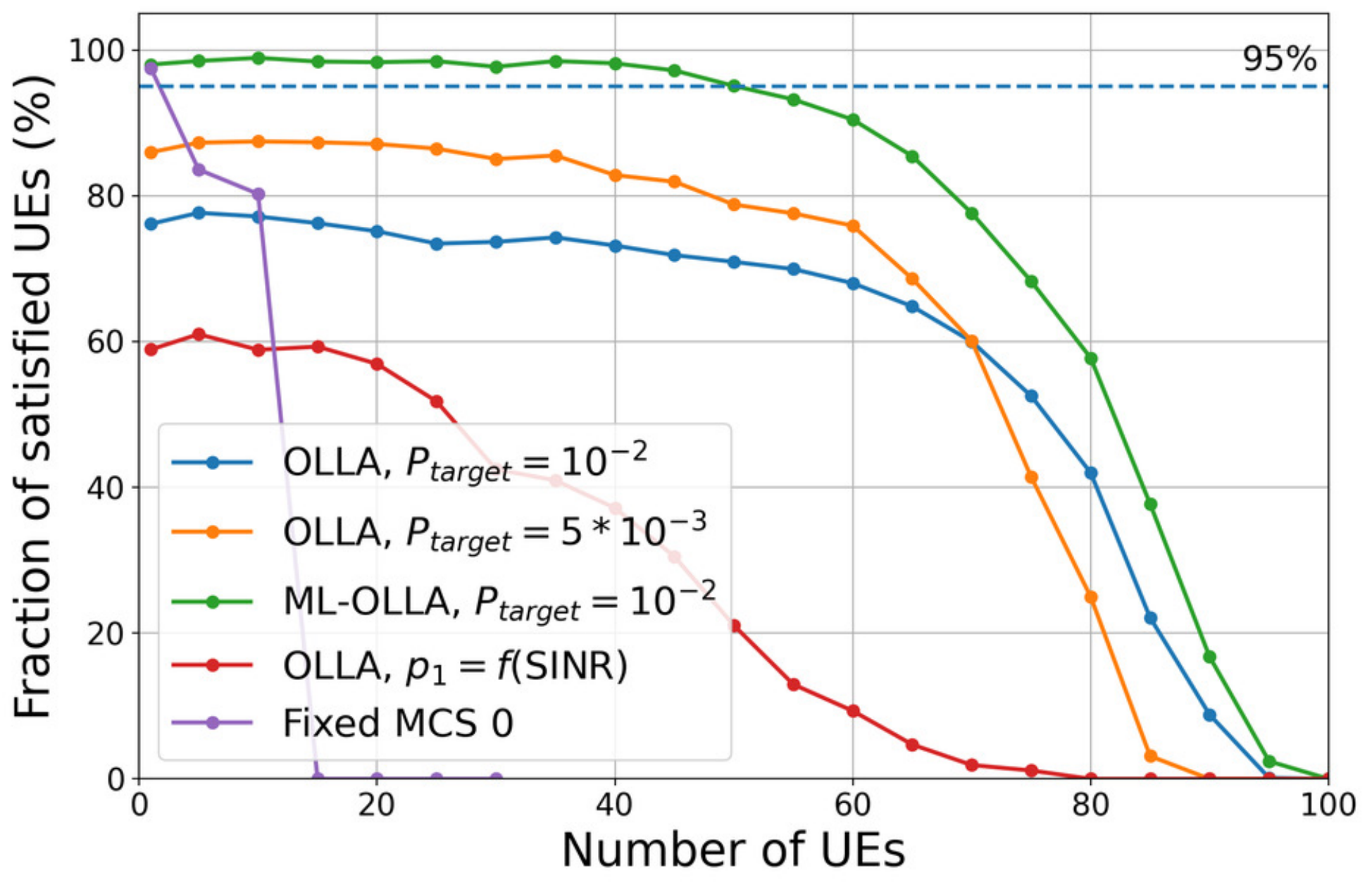}
    \caption{Fraction of satisfied UEs as a function of the number of UEs for the gNB on ground scenario, \(D^{QoS}=30\)~ms and \(PLR^{QoS}=10^{-2}\).}
    \label{fig:fraction_satisfied_ues}
  \end{minipage}\hfill
  \begin{minipage}[t]{0.485\linewidth}
    \centering
    \includegraphics[width=\linewidth]{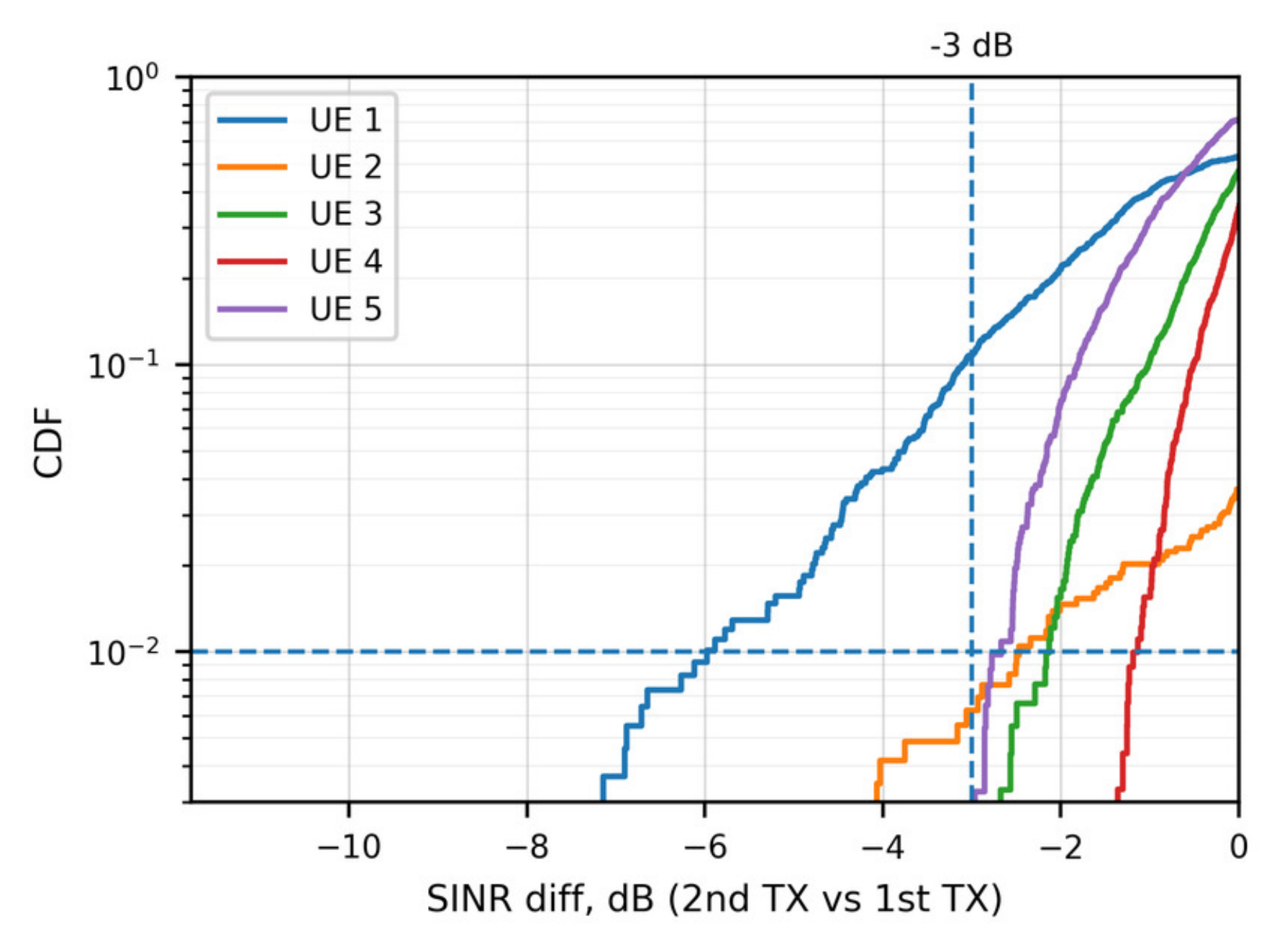}
    \caption{CDFs of SINR degradation from the 1st to 2nd transmission.}
    \label{fig:cdf_sinr_diff}
  \end{minipage}
\end{figure}

Fig.~\ref{fig:fraction_satisfied_ues} shows the results for $PLR^{QoS}=10^{-2}$ and $D^{QoS}=30$~ms. With these constraints and the considered satellite altitude, the gNB can perform $N=2$ transmission attempts. 
ML-OLLA is the only algorithm that remains above the \(95\%\) capacity threshold over a wide range of loads (i.e., number of UEs). The OLLA with $P_{target}=PLR^{QoS}$ satisfies less than 80\% UEs. Fig.~\ref{fig:cdf_sinr_diff} explains this phenomenon. For each UE, the CDF is built over packet transmissions by comparing the SINR available at the second transmission attempt minus SINR at the first one. The curves differ because the UEs have different positions, speeds, and channel conditions within the beam. For some UEs (e.g., see results for UE~1) the degradation of SINR between the first and second transmission attempts exceeds \(3\)~dB with probability higher than $PLR^{QoS}$. Since HARQ-CC can compensate only about \(3\)~dB, the second transmission attempt will be decoded with a lower probability than the first attempt, resulting in violation of the target PLR.
ML-OLLA mitigates this effect by using more conservative target BLER for retransmissions and therefore tolerates such channel drops. 

A way to overcome the problem mentioned above for OLLA is to set more conservative $P_{target}=5\cdot10^{-3}$. It increases the fraction of satisfied UEs but at the cost of higher channel resource consumption. We see a sharper degradation of the curve at high loads (i.e., when all UEs cannot transmit their packets in time because of the increased queuing delay).

\begin{figure}[!htbp]
  \centering
  \begin{minipage}[t]{0.485\linewidth}
    \centering
    \includegraphics[width=\linewidth]{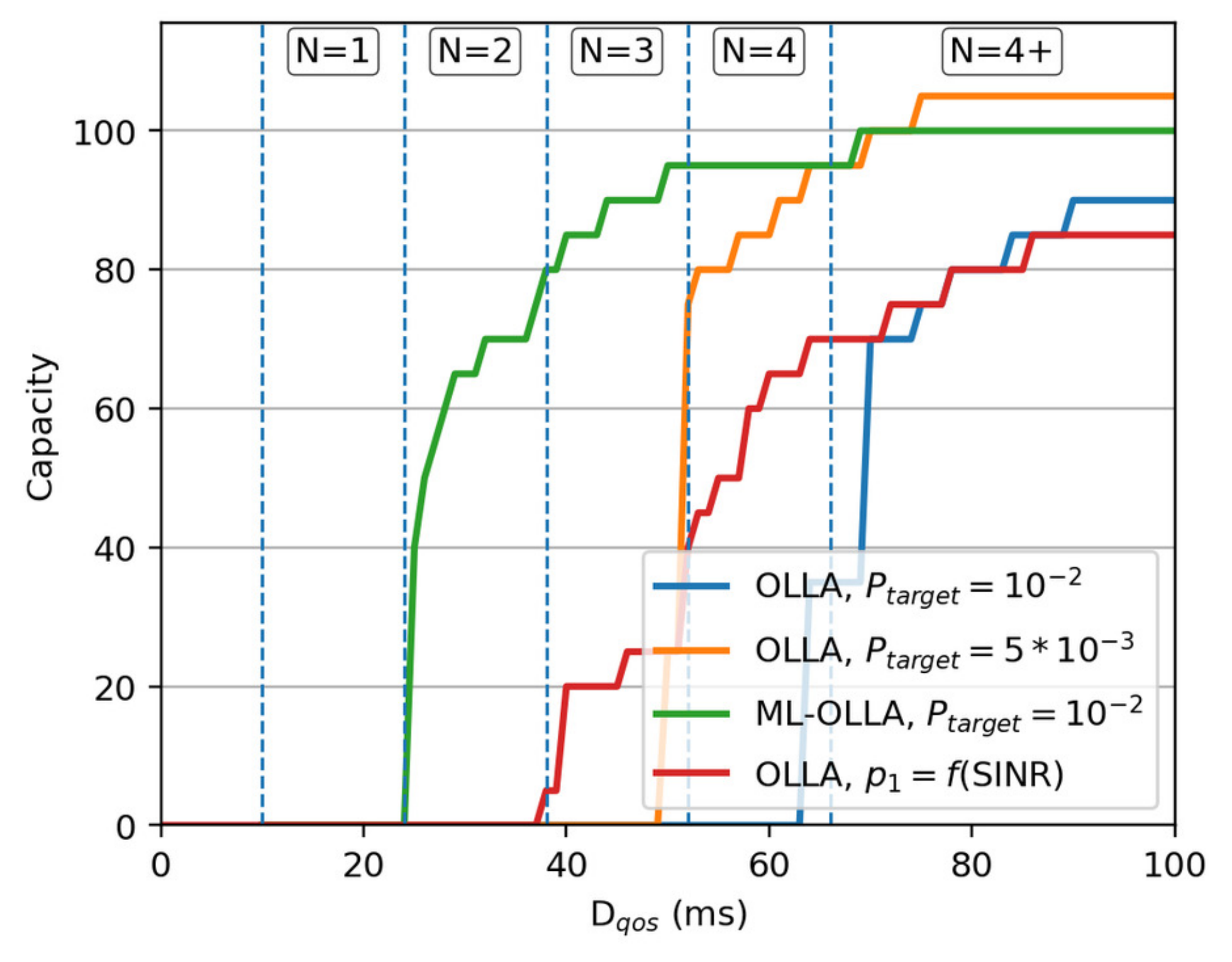}
    \caption{gNB on ground capacity versus $D^{QoS}$ for $PLR^{QoS}=10^{-2}$.}
    \label{fig:capacity_ground_1e2}
  \end{minipage}\hfill
  \begin{minipage}[t]{0.485\linewidth}
    \centering
    \includegraphics[width=\linewidth]{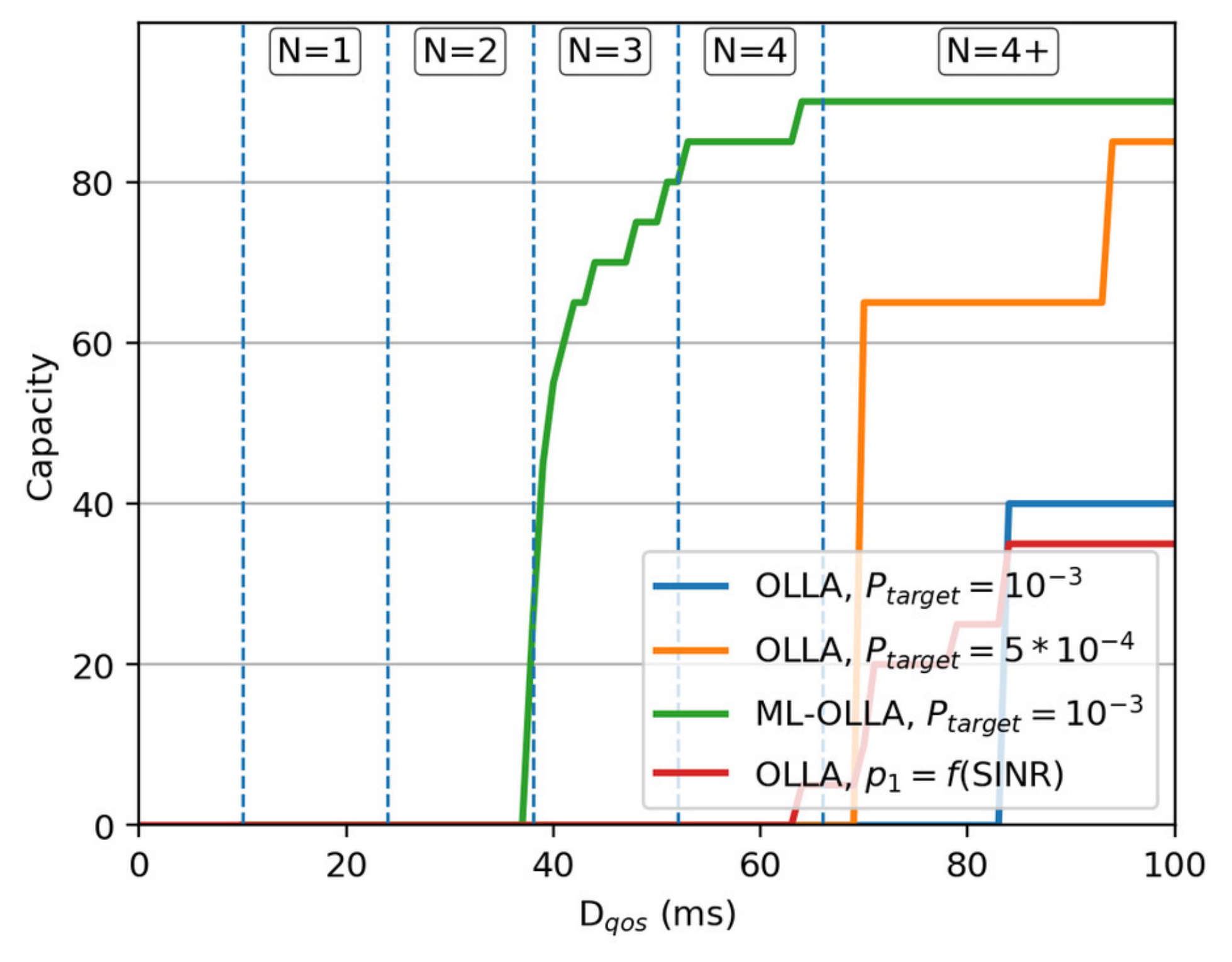}
    \caption{gNB on ground capacity versus $D^{QoS}$ for $PLR^{QoS}=10^{-3}$.}
    \label{fig:capacity_ground_1e3}
  \end{minipage}
\end{figure}

Fig.~\ref{fig:capacity_ground_1e2} shows the capacity as a function of $D^{QoS}$ for \(PLR^{QoS}=10^{-2}\). Among the baselines, only OLLA with $p_1=f(SINR)$ provides non-zero capacity in the interval $[40,50]$~ms, which corresponds to approximately three transmission attempts, and supports about 20 UEs. In contrast, ML-OLLA provides notable capacity at delay budgets down to about \(25\)~ms and supports about 80 UEs in $[40,50]$~ms, thus providing a fourfold gain.

The gap between ML-OLLA and baselines becomes even more pronounced for the stricter requirement \(PLR^{QoS}=10^{-3}\), as shown in Fig.~\ref{fig:capacity_ground_1e3}. ML-OLLA provides notable capacity at delay budgets down to roughly \(40\)~ms, whereas baselines achieve non-zero capacity for only delay budgets above \(60\)~ms. At \(D^{QoS}=70\)~ms, ML-OLLA provides nearly a \(30\%\) capacity gain. Note that stricter reliability requirements increase the minimum delay budget required by all solutions because higher reliability targets require compensating higher SINR degradation (see Fig.~\ref{fig:cdf_sinr_diff}), which eventually requires performing additional transmission attempts. 
\begin{figure}[!htbp]
  \centering
  \includegraphics[width=0.68\linewidth]{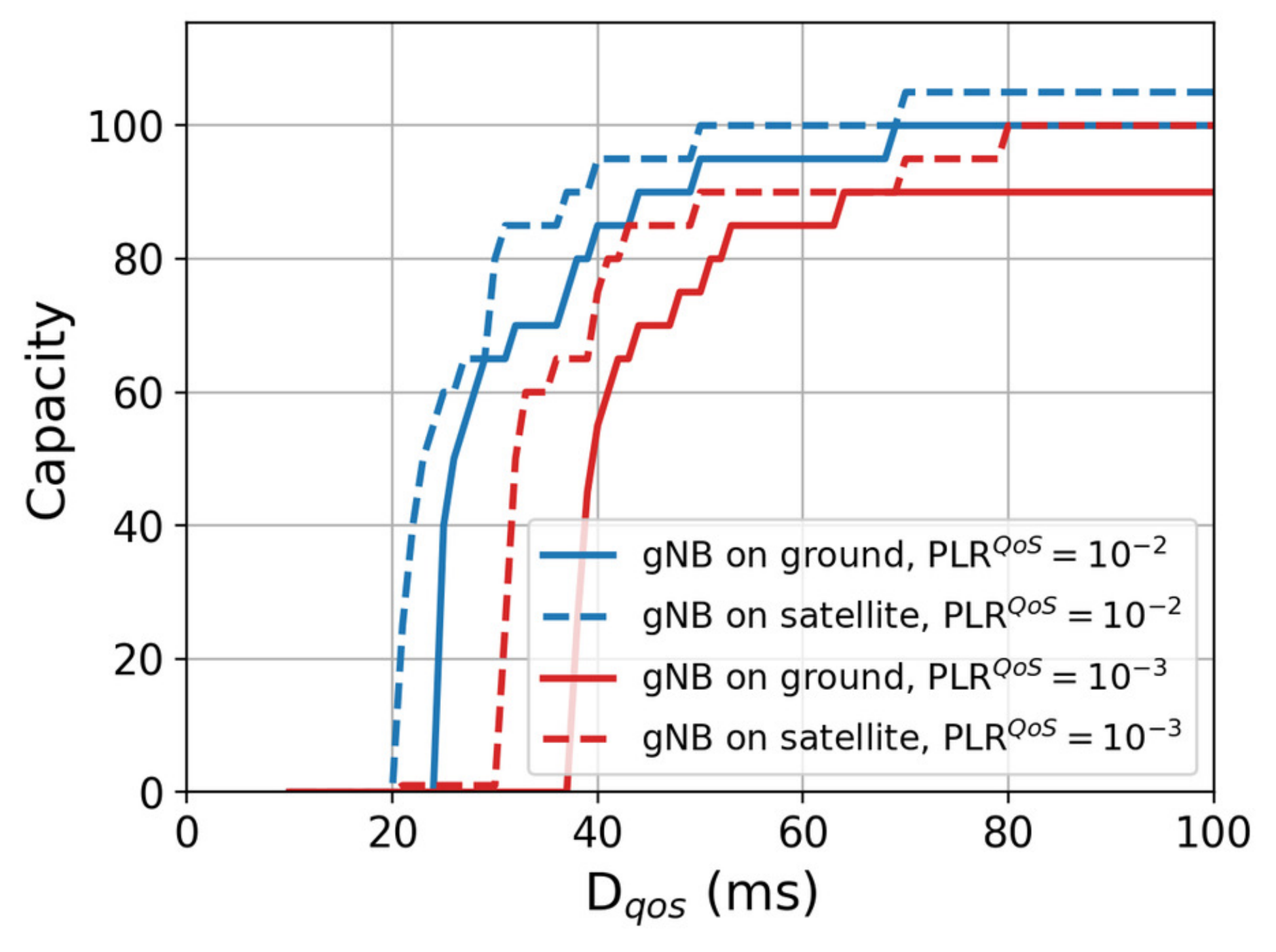}
  \caption{Capacity versus $D^{QoS}$ for different gNB deployments.}
  \label{fig:ground_vs_satellite}
\end{figure}

Finally, Fig.~\ref{fig:ground_vs_satellite} compares different gNB deployment options when the ML-OLLA algorithm is used. Placing the gNB on the satellite reduces the minimum supported delay budget from about \(25\) to \(20\)~ms for \(PLR^{QoS}=10^{-2}\), and from about \(40\) to \(30\)~ms for \(PLR^{QoS}=10^{-3}\). This gain is not a factor-of-two reduction because the gNB on satellite option also takes into account packet transmission delay over the feeder link. Thus, the first transmission attempt effectively takes the same time for both options, while HARQ retransmissions are performed twice as fast for the gNB on the satellite. In addition, the gNB on satellite deployment provides \(5\)--\(10\%\) capacity gain because CQI reports arrive on the ground gNB with a higher delay. Thus, the results show practical benefits of deploying gNB on the satellite for real-time applications with strict latency requirements.
\section{Conclusion}
\label{sec:conclusion}

In this paper, we developed the ML-OLLA algorithm for real-time traffic in 5G/6G NTN systems. The algorithm assigns a dedicated target BLER and an independent OLLA process to each transmission attempt, enabling more aggressive MCS selection for early attempts and more robust ones for retransmissions. Its data-driven parameter-selection method derives the per-attempt targets from the empirical distribution of recent CQI reports.

Numerical results obtained with ns-3 show that ML-OLLA with the proposed adaptive parameter selection method increases system capacity by up to four times compared with the baselines. Moreover, ML-OLLA reduces the minimum supported latency from 40 to 25~ms for $PLR^{QoS}=10^{-2}$ and from 60 to 40~ms for $PLR^{QoS}=10^{-3}$. These lower latency budgets extend the set of real-time applications that can be supported over LEO NTN systems. We also compare ground- and satellite-gNB deployments: placing the gNB on the satellite further reduces the minimum supported delay and provides a 5--10\% capacity gain in the evaluated scenario, because HARQ and CQI feedback are available earlier.

Future work will extend the ML-OLLA parameter-selection model to explicitly account for HARQ-CC gain. We also plan to take into account variations of MCS caused by OLLA and its influence on channel resource consumption.

\FloatBarrier
\bibliographystyle{IEEEtran}
\bibliography{references}

\end{document}